\documentclass[a4paper,12pt]{article} 

\usepackage[dvips]{graphicx}

\usepackage{bm}

\newcommand{\BBox}{\sqcup\!\!\!\!\sqcap }

\begin{document}

\begin{flushleft}
\Large

\textbf{Who runs fastest in an adaptive landscape: Sexual versus asexual reproduction\\}
\renewcommand{\thefootnote}{\fnsymbol{footnote}}
\vspace{.5cm}
\large
\textbf{Kerstin Holmstr{\"o}m and Henrik Jeldtoft Jensen\footnote[1]
{Author for correspondence (h.jensen@imperial.ac.uk; 
URL: http://www.ma.imperial.ac.uk/$\sim$hjjens)} \\}

\normalsize
\textit{Department of Mathematics, Imperial College London, South Kensington
campus,
 London SW7 2AZ, U.K. \\}
 
\end{flushleft}
\line(1,0){200}
\linespread{1}

\date{11.07.03}

\begin{abstract}
We compare the speed with which a sexual, respectively an asexual, population is able 
to respond to a biased selective pressure. Our model  focuses on the
Weismann hypothesis that the extra variation caused by crossing-over and
recombination during sexual reproduction allows a sexual population to
adapt faster. We find, however, that
the extra variation amongst the progeny produced during sexual reproduction 
for most model parameters is unable 
to overcome the effect that parents with a high individual fitness in general must mate 
with individuals of lower individual fitness resulting in a 
moderate reproductive fitness for the {\em pair}.
\end{abstract}

\noindent Short title: Sex and reproductive fitness.

\section{Introduction}
It is generally assumed that an important contribution to overcoming the evolutionary 
cost of sexual reproduction is the greater genetic variation among the progeny of sexual 
parents, caused by crossing-over and recombination, compared to asexually reproduced 
offspring. The greater  variation  in the sexual population is assumed to
produce more 
choice for natural selection and might thereby enable swifter adaptation. This is the 
hypothesis of Weismann \cite{burt}. Several recent experimental studies
have been interpreted as supporting Weismann's hypothesis \cite{Colegrave_1,Colegrave_2,Kaltz}.
These experiments observe a larger variation in the reproductive fitness
(rate of reproduction) when sexual reproduction is involved and this produces
a faster adaptation of the sexual population. Thus it might seem that Weismann's
suggestion has verified and fully understood.

However, we believe that there are still issues to be clarified concerning
the details of the mechanisms behind Weismann's hypothesis. The question
we discuss here is the following: It is evident that crossing-over and 
recombination may lead to greater genotypical variation amongst the progeny
but what matters for selection is the differential excess fitness (reproduction
rate) of sexual reproduction compared with asexual reproduction. How is
the greater genotypical variation amongst individuals able to lead to a larger reproduction
rate of the {\em pair of parents}? The essential point is that in a sexual
population the individual of highest fitness typically will have to mate with an
individual of lower fitness.

We use here a simple schematic mathematical model of evolution in genotype 
space and study the speed with which the population is able to climb a 
fitness gradient. Sexual individuals can mate at random with any other
individual of the same species, which we, according Mallet \cite{mall:spec}
genoptypic cluster definition, interpret as individuals of sufficient similar
genotype. Although our model is simple, it is nevertheless representative,
as far as testing Weismann's hypothesis, 
of any approach in which the fitness of the sexually reproducing pair is
obtained by combining measures of fitness of the two individuals involved
in the reproductive event. 
 We find that the sexual population generically exhibit
a greater spread in genotype space compared with the asexual population.
However, only in extreme cases does this greater variation in genotype space 
lead to a larger variation in the reproduction rate and thereby to
a superior adaptability for the sexual population.

\section{The Model}
{\it Genotype space:} We consider a population of individuals each characterised by 
their specific genotype determined by a position vector ${\bf S}$ in genotype space 
$\cal{S}$. For simplicity we model $\cal{S}$ by a two dimensional $L_x\times L_y$ 
grid, i.e. ${\bf S}=(S_x,S_y)$, where $S_x$ and $S_y$ are integers: $S_x=1,2,\dots,
L_x$ and $S_y=1,2,\dots
,L_y$, with periodic boundary conditions in the $y$-direction. We have
used $L_x=L_y=100$ for the study reported here.
The number of individuals 
at position ${\bf S}$ at time $t$ is given by $n({\bf S},t)$. 

{\it Reproduction rate:} The genotype determines directly the reproduction rate, or the 
reproductive fitness, $\phi({\bf S},t)$ in the following way
\begin{equation}
\phi({\bf S},t)=  {1\over 1+\exp[-W]}.
\label{phi(S)}
\end{equation} 
The statistical weight $W({\bf S},t)$ is given by 
\begin{equation}
W({\bf S},t) = \alpha S_x  -\mu N(t).
\label{W(S)}
\end{equation}
Here the first term represents a linear increase ($\alpha>0$) in the reproduction rate 
along the $x$-direction in genotype space. In the second 
term $N(t)=\sum_{\bf S}n({\bf S},t)$ represents the total number of individuals 
at time $t$, and $\mu$ determines the carrying capacity of the system. For
fixed $N(t)$ the fitness 
landscape corresponding to $W({\bf S},t)$ in Eq.~(\ref{W(S)}) is a smooth upwards 
tilted plane. Assume for a moment that the entire population is placed on one 
location in genotype space, say ${\bf
S}^0$, and assume further more that adaptation cannot take place 
(zero mutation rate). A stationary population size is then determined by the balance
between the reproduction [$N\rightarrow N+N\phi({\bf S})$] and ensuing annihilation of individuals 
[$(1+\phi({\bf S}))N\rightarrow (1+\phi({\bf S}))N(1-p_{kill}$)], leading to
$\phi({\bf S})=p_{kill}/(1-p_{kill})$. From this expression we find the
 population size, for a given carrying capacity $\mu$, by use of Eq. (\ref{phi(S)}) and (\ref{W(S)}), 
\begin{equation}
N = {1\over\mu}[\alpha S^0_x- \ln\left({\phi\over 1-\phi}\right)]={1\over\mu}[\alpha
S^0_x-\ln\left({p_{kill}\over 1-2p_{kill}}\right)].
\label{N_of_x}
\end{equation}
When mutations are allowed the population can adapt (done by increasing
the $S_x$ component of the position in genotype space) to the environment
and thereby increase the population size for a given value of the carrying capacity
parameter $\mu$.
We have used the following parameter values: $\alpha = \ln 3/45$ and $\mu = \ln 3/450$
and $p_{kill}=0.2$
which leads to the linear dependence $\langle N\rangle =10\langle S_x\rangle + 450$ 
of the average number of individuals $\langle N\rangle$ on the population
averaged x-coordinate $\langle S_x\rangle$ in genotype space.

In order to be able to represent a more rough landscape we choose a
fraction of genotype positions ${\bf S}_h$ at random and denote them fitness
holes. The $y$-direction in the genotype space is essential in this case
as this extra dimension allows a population to bypass a fitness hole.
These sites are 
infinitely unfit corresponding to $\phi({\bf S}_h)=0$, i.e. individuals with a 
genotypical composition equal to one of the ${\bf S}_h$ locations cannot produce offspring. 
The fitness landscape produced by this procedure resembles  Gavrilets'  notion of
the holey 
fitness landscape \cite{gavrilets,glv_1,glv_2}. 

{\it Time step :} 
 The system is simultaneously updated.  First all individuals are allowed to reproduce 
 by the relevant probability to be described below. After reproduction annihilation 
 occurs, during which individuals are removed from the system with probability $p_{kill}$. 
 This probability is constant in time and equal for all genotypes.

{\it Asexual reproduction:} Let us now describe how reproduction is represented in the 
model. We begin with the asexual case. 
All $N(t)$ individuals are allowed to produce one offspring with the genotype
specific probability $p_{off}({\bf S})=\phi({\bf S},t)$. The offspring will be of
the same genotype ${\bf S}$ except when mutations occur, which happens  
with a constant probability $p_{mut}$, to be thought of as the genomic
mutation rate per reproduction event. The mutant is placed at random on one of 
the four nearest neighbour sites.

{\it Sexual reproduction:} An individual can mate with any other individual
if the two are sufficiently similar to belong to the same species. We take
this to mean that two matting individuals must share a certain degree of
genotypical similarity, cf. Mallet's  genotypic cluster species 
definition Mallet \cite{mall:spec}. We consider two individuals
to be of the same species if they belong to genotypes which are separated
by less than a distance $d_{max}$ in genotype space.

\begin{figure}[h]
\includegraphics{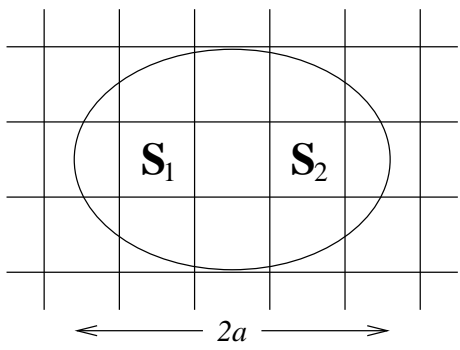}
\caption{Ellipse with $r=1.4$. $S_1$ and $S_2$ denote the sites
of the two parents. The distance $a$ is the major axis defined in the text. }
\label{Ellipse_fig}
\end{figure}

First a mate of genotype ${\bf S}_{mate}$ is found for each individual $i$
of genotype ${\bf S}_i$ for $i=1,2,\dots,N(t)$, This is done by randomly
choosing the mate amongst all the individuals present on genotypical positions less than a 
distance $d_{max}$ away from ${\bf S}_i$. In this way we ensure that only
individuals belonging to the same species mate and that there is no preferential mating
within the species. 
Each of the $N(t)$ {\em  pairs} will 
produce offspring with the respective probability 
\begin{equation}
p_{off}=\sqrt{\phi({\bf S}_i)\phi({\bf S}_{mate})}.
\label{p_off}
\end{equation} 
The  variation amongst the offspring induced by crossing-over and recombination is 
represented by placing the offspring on a site ${\bf S}_{off}$ selected randomly 
with uniform probability within an ellipse with eccentricity $e$ and major half axis 
of length $a=[1/2+|{\bf S}_1-{\bf S}_2|+1/2]/2e$ placed symmetrically around ${\bf S}_1$ 
and ${\bf S}_2$, see Fig.~\ref{Ellipse_fig}. The likelihood with which the offspring are allowed 
to end up {\em outside} the region in genotype space {\em between} the two parents is 
characterised by what we call the recombination factor $r=1/e$. The larger the value 
of $r$ the more likely it is to find the offspring to the right (or to the left) of 
both parents, see Fig.~\ref{Ellipse_fig}. In the special case where both
parents are of identical genotype, i.e. ${\bf S}_i={\bf S}_{mate}$, the
offspring is given the same genotype as the parents, i.e. ${\bf S}_{off}={\bf
S}_i$.

Mutations can occur with the
same probability $p_{mut}$ as in the asexual case. When a mutation occur 
the offspring is moved to one of the nearest neighbour sites of the 
position ${\bf S}_{off}$.

We emphasise that the specific function in Eq. (\ref{p_off}) is inessential.
We obtain qualitative identical results when we instead of Eq. (\ref{p_off})
use other functional forms for $p_{off}$  which satisfies the seemingly reasonable
requirement that $p_{off}$ is a number between the individual fitnesses 
$\phi({\bf S}_i)$ and $\phi({\bf S}_{mate})$: for example one can replace 
Eq. (\ref{p_off}) by $p_{off}=[\phi({\bf S}_i)+\phi({\bf S}_{mate}]/2$. 

It is also important to point out that although the population
grows monotonically in size as it moves to the right in genotype space, the average
reproduction probability $\langle p_{off}({\bf S}) \rangle$  remains constant.
The reason for this is that the increase in fitness caused by the increase
in $\alpha S_x$ in Eq.~(\ref{W(S)}) is compensated by the carrying capacity term
$-\mu N(t)$ in this equation.

\section{Results}
In order to compare the efficiency with which the two different types of reproduction 
respond to a selective pressure, we compare the evolution under the same conditions 
(i.e. same $\alpha$ and $\mu$ in Eq. (\ref{W(S)})) of a solely sexually reproducing 
population with a solely asexually reproducing population. 
In both the asexual and sexual case five different genotypes are 
initiated with 100 individuals each. They are chosen
so that the average $S_x$-coordinate $\langle S_x\rangle =5$. The five
positions are ${\bf S}=(4,L_y/2),(5,L_y/2),(6,L_y/2),(5,L_y/2+1),(5,L_y/2-1)$. 
This gives an initial reproduction 
probability for all the 500 individuals of about $p_{off}=0.252$. 
The exact number of starting genotypes does not 
affect the nature of the movement through genotype space. 

We monitor the time it takes the population's centre of mass $\langle {\bf S}\rangle$
to move in genotype space from $\langle {\bf S}\rangle=5$ to $\langle {\bf S}\rangle=85$.
During this motion we find that $\langle {\bf S}\rangle$ to very a good approximation
increases linearly with time.

\begin{figure}[h]
\includegraphics{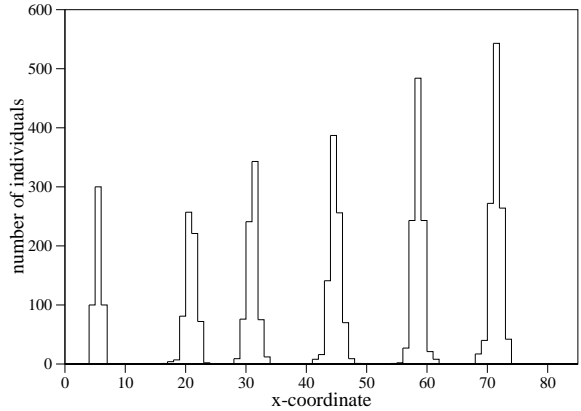}
\caption{The peak in the population density moving to the
right along the $S_x$ axis in genotype space. The data is for a sexually
reproducing population with $p_{mut}=0.01$, $d_{max}=\sqrt{2}$ and $r=1.27$.
The time between each small peak is equal to 6000 time steps.}
\label{peaks}
\end{figure}

 Under the evolutionary dynamics the population moves to the right.
 Although the 
 dynamics is diffusive in nature the population remains confined to a rather narrow 
 region along the $S_x$ and $S_y$ axes while the population gradually moves to lager $S_x$ values. This is 
 shown as the peaks in $n({\bf S},t)$ in Fig. \ref{peaks}. The dispersive action of 
 the diffusion in genotype space produced by the mutations, and the crossing-over 
 effect in the sexual case, is counteracted by the effect of $-\mu N(t)$ in Eq. 
 (\ref{W(S)}). The reproduction probability $p_{off}({\bf S})$ of individuals left behind, 
 at genotype positions with small $S_x$ values, will decrease with time because the term 
 $-\mu N(t)$ increases as a result the reproductive activity of individuals with 
 higher $S_x$ values. As soon as the reproductive probability of a site ${\bf S}$ 
 becomes permanently smaller than the killing probability, $p_{off}({\bf S},t)<p_{kill}$, 
 the population on that site is bound to go extinct.

\begin{figure}[h]
\includegraphics{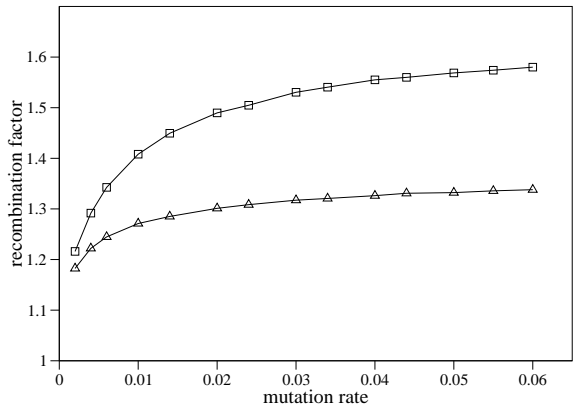}
\caption{The competition between sexual and asexual
reproduction for two different values
of $d_{max}$. The signatures are: $\triangle$ corresponds to $d_{max}=\sqrt{2}$ and $\BBox$
to $d_{max}=\sqrt{5}$. For a given $d_{max}$, sexual reproduction most 
efficiently response to the selective pressure above the relevant curve, below the curve asexual 
reproduction is superior. }
\label{diagram1}
\end{figure}

We now summarise the behaviour of the model. First we consider the case without any 
fitness holes in genotype space. The sexually reproducing population is 
found only to adapt more rapidly than the asexually reproducing population for substantial 
variation amongst the progeny or low mutation rate.  This is illustrated in Fig. 
\ref{diagram1}. For a given mutation rate, 
the data points indicate the threshold value, $r_{thr}$, of the recombination factor 
where, for increasing $r$, sexual reproduction becomes more effective than asexual
 reproduction. It is remarkable that a sexually reproducing population
 is only able to adapt 
 faster than an asexually reproducing population if the recombination factor $r$ is significantly 
 larger than 1. Further more we note that the threshold value of $r_{thr}$ is 
 slowly increasing with increasing mutation rate. This is, of course, to be expected
 since the asexual population is able to adapt faster with increasing mutation rate.

It is clearly difficult to relate in a detailed way the mutation rate $p_{mut}$
in this model to real biological mutation rates. Fig.~\ref{diagram1} shows
that this is not an essential problem for the interpretation of our results
since the same qualitative behaviour is found for all values of the mutation
rate $p_{mut}$. Namely, recombination factors $r$ significantly in excess of 1 is needed
for all values of the mutation rate to make the sexual population
adapt faster than the asexual population.

\begin{figure}[h]
\includegraphics{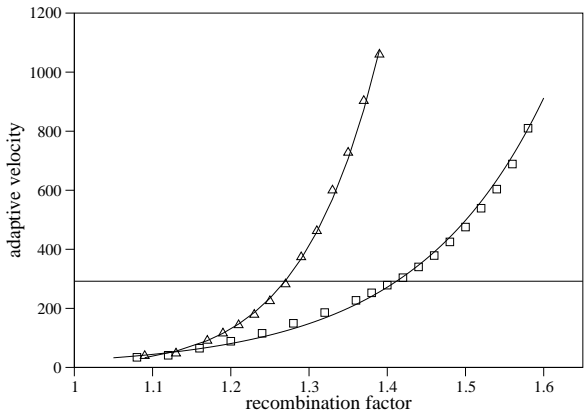}
\caption{The adaptive velocity of a sexually reproducing population for different 
values of the recombination factor and different values of $d_{max}= \sqrt{2}$,
$\triangle$,  and $\sqrt{5}$, $\BBox$, respectively. The straight 
line represent the velocity ($229\times 10^{-5}$ square/time step)
of an asexually reproducing population for the same value
of the mutation rate $p_{mut}=0.01$. The two intersection points mark the
minimum recombination factors at which sexual reproduction is superior
to asexual ditto.}
\label{diagram2}
\end{figure}

The results in Fig. \ref{diagram1} can be understood in terms of how the velocity of 
the centre of mass of the population depends on the recombination factor for the 
two types of reproduction. This is illustrated in Fig. \ref{diagram2}.
By construction the velocity of the asexually reproducing population is independent of the 
recombination factor. The velocity of a sexually reproducing population increases 
rapidly with increasing recombination factor, but decreases with increasing maximum 
distance $d_{max}$ between two reproducing parents. This finding may be
understood in the following terms. Consider a reproduction event involving
a {\em most} fit individual, i.e. an individual on one of the positions,
say ${\bf S^*}$, in genotype space with the highest value of the $S_x$-coordinate among
the genotypical positions occupied at the present instant in time. I.e.,
the positions in genotype space with $S_x>S^*_X$ are all unoccupied.
An increase in the recombination factor leads to an increase in the range
of possible $S_x$-values of the  offspring's genotype and thereby assists the
displacement of the population to large values of $S_x$. Consider again
this {\em most} fit individual ${\bf S}^*$. The average reproduction rate of 
${\bf S}^*$ is obtained from Eq. (\ref{p_off}) by averaging over all potential 
individuals present within the distance $d_{max}$. The larger the value of
$d_{max}$ the smaller the $S_x$ values of the possible mates. Small values
of $S_x$ lead to small reproduction values according to Eqs. (\ref{phi(S)}),
(\ref{W(S)}) and (\ref{p_off}).

The results discussed in this section are all for the case where no fitness
holes are present. In this case the simulations can readily
be compared with the iterative master equation for the population density
in genotype space $n({\bf S},t)$.  These are cumbersome to write down but
can easily be solve by numerical methods. In this way we have confirmed
the ensemble averaged results described above.

\section{The effects of holes}
In reality many totally unviable genotypes exist. We include this in the
model by introducing a 
random selection of genotype positions which are unable to reproduce. Individuals can 
arrive at one of these positions as a result of mutation from a nearby site. 
Individuals on such a site, say ${\bf S}_h$, may be selected for reproduction attempts. 
In both the sexual and asexual case this reproduction attempt can, however, not lead to 
offspring since $\phi({\bf S}_h)=0$. The individuals on the trap sites leads to an 
overall fall in the reproduction rate on other sites through their contribute to 
the $-\mu N(t)$ term in $W$, see Eq. \ref{W(S)}. This is only natural as all individuals 
existing at a given moment in time represent a demand on the carrying capacity of the 
system. One might expect that the sexually reproducing population would be more likely 
to find a path through this rugged fitness landscape than an asexually reproducing 
population. After all, the sexually reproducing population may jump over holes in the 
fitness landscape when $d_{max}\geq 2$. In fact, as soon as $d_{max}\geq
\sqrt{2}$ the sexual population can pass to next nearest neighbour sites
of already occupied positions in genotype space,
whereas the asexual offspring can at most move to a nearest neighbour
site when a mutation occur. This enables the sexual population to follow
paths through the rugged fitness landscape which are inaccessible to the
asexual population.

\begin{figure}[h]
\includegraphics{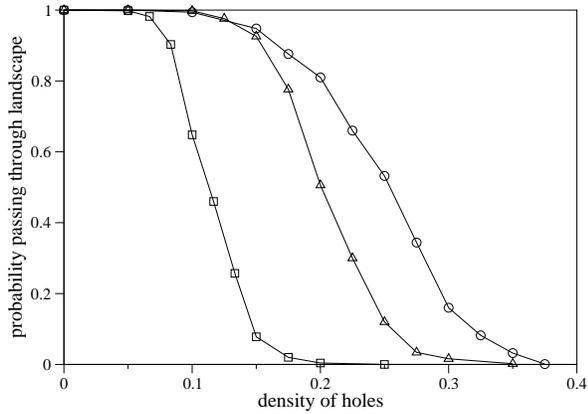}
\caption{The fraction of the population that on average is able to pass through 
a rough fitness landscape with traps.The signatures are as follows: $\circ$ asexual
population, $\triangle$ sexual population with $d_{max}=\sqrt{2}$ and $r=1,27$,
$\BBox$ sexual population with $d_{max}=\sqrt{5}$ and $r=1.41$. In all
cases $p_{mut}=0.01$ and each data is averaged over 500 realisations.}
\label{diagram3}
\end{figure}

In Fig. \ref{diagram3} we show as function of the density of holes 
the probability that a population is able to find 
an adaptive path through the rough fitness landscape in the course of a fixed number 
of time steps $T=10^6$ for mutation rate $p_{mut}=0.01$.
We present results for two cases: A) $d_{max}=\sqrt{2}$ with $r=1.270$
and B) $d_{max}=\sqrt{5}$ with $r=1.408$. In both cases $r$ is calibrated 
such that, in the absence of holes, the sexual population moves with the same velocity 
through genotype space, as the asexual population with $p_{mut}=0.01$ does.

The simulations clearly show that, once holes are present, the asexual 
reproduction in general is better suited to move through the rough fitness landscape, 
and that large values of $d_{max}$, rather than allowing the sexual population to more 
easily find a way around the traps, has the opposite effect. This behaviour 
is explained by the fact that the sexually reproducing population more
often place offspring in the holes than the asexual population does.

\section{Discussion}
\begin{figure}[h]
\includegraphics{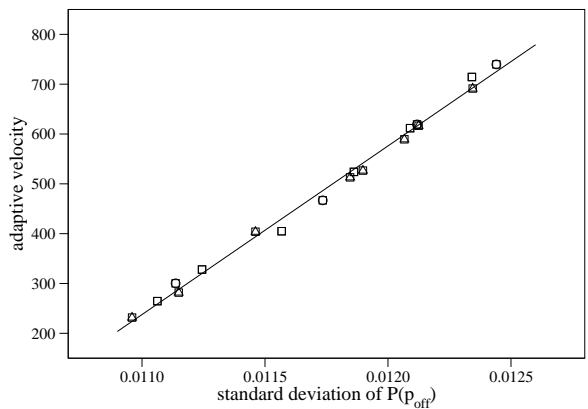}
\caption{Average adaptation velocity as function of the standard deviation
of the distribution of reproduction rates. The data are for $p_{mut}=0.01$
and $0.03$, $d_{max}=\sqrt{2}$ and $\sqrt{5}$ for a range of different
recombination factors $r$. }
\label{width}
\end{figure}

It is interesting to relate our results to the width of the population along the x-axis in
genotype space and to the width of the distribution of the reproduction probabilities. We find
that the population that is able to adapt most rapidly always has the {\em
broadest} distribution of reproduction probabilities $p_{off}$. This is
entirely consistent with the Weismann's idea that greater variation enables
adaptation to occur faster because natural selection has more choice to
act upon \cite{burt}. What is not consistent with Weismann's hypothesis
is that sexual reproduction does not always lead to the larger variation.
In Fig. \ref{width} we show the average velocity of adaptation of the sexual 
and asexual population as function of the standard variation $\sigma_{p_{off}}$ of the reproduction 
rate of the respective population for many different combinations of control parameters.

We notice that adaptive velocity and $\sigma_{p_{off}}$ are linearly related
and that  only for very substantial
recombination factors is the sexually reproducing population able to adapt
faster than the asexual. When this happens the sexual population
does have the greatest variation in $p_{off}$.
 Recombination factors $r$ much larger than one seems of little biological 
 relevance since $r>1$ frequently
leads to offspring which may be reproductively isolated from one of its parents in
the sense that the distance to one of the parents may exceed $d_{max}$.
Thus, in the model considered here a sexual population
is unable to adapt faster than an equivalent asexual population except
for, what seems to be, biologically extreme choices of parameters.

The reason we find that Weismann's hypothesis is unable to explain the
superiority of sexual reproduction is not that the idea that greater variation
leads to faster adaptation is wrong. The problem is that sexual
reproduction events involving the most fit individuals will typically involve
less fit mates. This hinders the individuals with the best adapted genotype to spearhead
the adaptation of the sexual population. 

Of course the Weismann hypothesis is not the only reason why sexual reproduction
might be evolutionary superior to asexual reproduction. Kondrashov 
discuss a large number of different hypotheses and emphasis in particular the ability 
of recombination
to eliminate deleterious mutations from the genome \cite{Kondrashov}. Perhaps
no single general mechanism can be identified as the most important for the evolution
and maintenance of sex (see e.g. \cite{hurst}) in which case the explanation
may have to rely on case specific mechanisms (see e.g. \cite{peck,doncaster}).

We have studied Weismann's hypothesis in it simplest form and have been
forced to conclude that within the framework of individual genotypical
fitnesses combined, in the sexual case, to give the fitness of the 
reproducing pair, 
the extra variation in genotype space of the sexual population is in realistic 
cases not able to produce the needed excess in reproductive fitness
to make the sexual population able to out compete the asexual population.
The sexual population's greater variation of genotypes 
tends to be neutralized by the necessity of
the most fit individual to mate with individuals of lower fitness.

We are greatful to A. Burt and C. Godfray for comments. We thank
K. Dahlstedt and G. Pruessner for assistance with prrgramming and
P. Anderson for lingusitic advice.

\bibliographystyle{apalike}

\bibliography{evolution}

\begin{thebibliography}{}

\bibitem[Burt, 2000]{burt}
Burt, A. (2000).
\newblock Perspective: Sex, recombination, and the efficacy of selection -- was
  {W}eismann right?
\newblock {\em Evolution}, 54:337--351.

\bibitem[Colegrave, 2002]{Colegrave_1}
Colegrave, N. (2002).
\newblock Sex releases the speed limit on evolution.
\newblock {\em Nature}, 420:664--666.

\bibitem[Colegrave et~al., 2002]{Colegrave_2}
Colegrave, N., Klatz, O., and Bell, G. (2002).
\newblock The ecology and genetics of fitness in {\it {c}lamydomonas.} {VIII}.
  the dynamics of adaptation to novel environment after a single episode of
  sex.
\newblock {\em Evolution}, 56:14--21.

\bibitem[Doncaster et~al., 2000]{doncaster}
Doncaster, C., Pound, G., and Cox, S. (2000).
\newblock The ecological cost of sex.
\newblock {\em Nature}, 404:281--285.

\bibitem[Garvilets, 1999]{gavrilets}
Garvilets, S. (1999).
\newblock A dynamical theory of speciation on holey adaptive landscapes.
\newblock {\em Am. Nat.}, 154:1--22.

\bibitem[Gavrilets et~al., 1998]{glv_1}
Gavrilets, S., Li, H., and Vose, M.~D. (1998).
\newblock Rapid paraptric speciation on holey adaptive landscapes.
\newblock {\em Proc. R. Soc. Lond. B}, 265:1483--1489.

\bibitem[Gavrilets et~al., 2000]{glv_2}
Gavrilets, S., Li, H., and Vose, M.~D. (2000).
\newblock Patterns of paraptric speciation.
\newblock {\em Evolution}, 54:1126--1134.

\bibitem[Hurst and Peck, 1996]{hurst}
Hurst, L. and Peck, J. (1996).
\newblock Recent advances in understanding of the evolution and maintenance of
  sex.
\newblock {\em Trends Ecol. Evol.}, 11:46--52.

\bibitem[Kaltz and Bell, 2002]{Kaltz}
Kaltz, O. and Bell, G. (2002).
\newblock The ecology and enetics of fitness in {\it {c}hlamydomonas}. {XII}.
  repeated sexual episodes increases rate of adaptation to novel environments.
\newblock {\em Evolution}, 56:1743--1753.

\bibitem[Kondrashov, 1993]{Kondrashov}
Kondrashov, A.~S. (1993).
\newblock Classification of hypotheses on the advantage of amphimixis.
\newblock {\em J. Heredity}, 84:372--387.

\bibitem[Mallet, 1995]{mall:spec}
Mallet, J. (1995).
\newblock {A} species definition for the {M}odern {S}ynthesis.
\newblock {\em Trends Ecol. Evol.}, 10(7):294--299.

\bibitem[Peck and Waxman, 2000]{peck}
Peck, J. and Waxman, D. (2000).
\newblock Mutation and sex in a competitive world.
\newblock {\em Nature}, 406:399--404.

\end{thebibliography}
\newpage

\end{document}